# Characterization of the Core-Shell Nanoparticles Formed as Soluble Hydrogen-Bonding Interpolymer Complexes at low pH

Maria Sotiropoulou[1], Frederic Bossard[2], Eric Balnois[3], Julian Oberdisse[4] and Georgios Staikos[1*]

[1]*Department of Chemical Engineering, University of Patras, GR-26504 Patras, Greece*
[2]*Institute of Chemical Engineering and High Temperature Chemical Processes, FORTH/ICE-HT, P.O. Box 1414, 26504 Patras, Greece*[1]
[3]*Laboratoire Polymères, Propriétés aux Interfaces et Composites (L2PIC), Université de Bretagne Sud, Rue de Saint Maudé, BP 92116, 56321 Lorient, France.*
[4]*Laboratoire des Colloïdes, Verres et Nanomatériaux, UMR CNRS/UM2, Université Montpellier II, F-34095 Montpellier, France and Laboratoire Léon Brillouin CEA/CNRS, CEA Saclay, 91191 Gif sur Yvette, France*

[1] *Present address: Laboratoire de Rheologie, UMR 5520, Université Joseph Fournier, 1301, rue de la piscine, BP 53, 38041 Grenoble Cedex 9, France.*

[*] To whom correspondence should be addressed. E-mail: staikos@chemeng.upatras.gr

ABSTRACT: The formation of soluble hydrogen-bonding interpolymer complexes between poly(acrylic acid) (PAA) and poly(acrylic acid-*co*-2-acrylamido-2-methyl-1-propane sulfonic acid)-*graft*-poly(*N*, *N* dimethylacrylamide) (P(AA-*co*-AMPSA)-*g*-PDMAM) at pH = 2.0 was studied. A viscometric study showed that in semidilute solution a physical gel is formed, due to the interconnection of the anionic P(AA-*co*-AMPSA) backbone of the graft copolymer, in a transient network, by means of the complexes formed between the PDMAM side chains of the graft copolymer and PAA. Dynamic and static light scattering measurements, in conjunction with small angle neutron scattering measurements, suggest the formation of core-shell colloidal nanoparticles in dilute solution, comprised by an insoluble PAA/PDMAM core surrounded by an anionic P(AA-*co*-AMPSA) corona. Even if larger clusters are formed in semidilute solution, the size of the insoluble core remains practically stable. Atomic force microscopy performed under ambient conditions, reveal that the particles collapse and flatten upon deposition on a substrate, with dimensions close to the ones of the dry hydrophobic core.

**Introduction**

When weak polyacids, such as poly(acrylic acid) (PAA) or poly(methacrylic acid) (PMAA), and proton acceptor polymers, such as polyethyleneoxide (PEO), or polyacrylamides, are mixed in solution at pH lower than 3–4, a associative phase separation takes place,[1, 2, 3, 4, 5, 6, 7] as a result of the formation of hydrogen-bonding interpolymer complexes (IPCs). A considerable amount of work on such hydrogen-bonding IPCs has been presented in two reviews.[8, 9] The potential applications of these IPCs to various fields, such as drug delivery formulations,[10, 11, 12] biomaterials,[13] emulsifiers,[14] and membrane and separation technology,[15, 16] has further stimulated the research interest in this field.

In order to extend the solubility of the hydrogen-bonding IPCs in the low pH region, some efforts have recently been undertaken.[17, 18] In one of them, an anionically charged graft copolymer, poly(acrylic acid-*co*-2-acrylamido-2-methyl-1-propane sulfonic acid)-*g*-poly (*N,N*-dimethylacrylamide) (P(AA-*co*-AMPSA)-*g*-PDMAM), has been synthesized by grafting poly (*N,N*-dimethylacrylamide) (PDMAM) chains onto an acrylic acid-*co*-2-acrylamido-2-methy-1-propane sulfonic acid copolymer (P(AA-*co*-AMPSA)) backbone. PDMAM is a water-soluble polymer with important proton acceptor properties, forming hydrogen-bonding IPCs with PAA,[19, 20] which precipitate out from water even at pH values as high as 3.75.[17] When these graft copolymers are mixed with PAA in a low pH (pH < 3.75) aqueous solution, hydrogen-bonding IPCs between the PDMAM side chains and PAA are formed. Nevertheless, the presence of the negatively charged AMPSA units in the graft copolymer backbone prevents their precipitation.[17] Moreover, rheological measurements in a semi-dilute solution have shown a gel-like behaviour in this low pH region.[21] This behaviour has been attributed to the interconnection of the negatively charged backbone chains of the graft copolymer by means of the hydrogen-bonding interpolymer complexes formed between the PDMAM side chains of the graft copolymer and the PAA chains.

Small-angle neutron scattering (SANS) measurements, already used to show the formation of dense hydrogen-bonding IPCs between PEO and partially neutralized (3 – 9%) PMAA,[22] were also used to study the microstructure of the above-mentioned electrostatically stabilized colloidal system in $D_2O$.[23] Core-shell nanoparticles comprised by an insoluble hydrogen-bonding IPC core and a hydrophilic negatively charged corona surrounding it, was supposed to be formed. The formation of similar colloidal complexes has been also observed as a result of the interaction of polyelectrolyte-neutral block copolymers or of comb-type polyelectrolytes with oppositely charged synthetic or biological macromolecules[24, 25, 26, 27, 28] and surfactants.[29,30, 31, 32, 33] However, there is a major difference in our case. The colloidal nanoparticles formed are pH sensitive as they are formed at low pH and dissociate at pH > 3.75.[17]

In this work we have proceeded to a thorough study of the interactions between a P(AA-co-AMPSA)-g-PDMAM graft copolymer, containing 48 wt % of PDMAM, shortly designated as G48, and PAA, at pH = 2.0, in a broad concentration region, ranging in the dilute and the semidilute regime. In such a low pH insoluble IPCs are formed[17] as a result of successive hydrogen bonds between the carboxylic groups of PAA and the amide groups of PDMAM.[6, 34] Nevertheless, the particles formed do not precipitate but remain in a colloidal form in the solution, due to the anionic backbone of the graft copolymer. Rheology measurements were used for the determination of a critical concentration, $c^*$, over which gel formation takes place. Dynamic light scattering (DLS) and SANS measurements indicated the formation of core-shell nanoparticles, transformed to bigger clusters as concentration increased above $c^*$, with their cores remaining unchanged. Static light scattering in dilute solution was used to determine the molecular weight of the isolated nanoparticles, and atomic force microscopy (AFM) to estimate their size after evaporation of the solvent.

**Experimental Section**

**Materials.** A sample of PAA (Polysciences), with a nominal molecular weight of 9.0 x $10^4$ Da, was dissolved in a 0.01N HCl solution, dialyzed against water through a cellulose membrane with a molecular weight cutoff equal to 12 kDa (Sigma), and finally obtained by freeze-drying.

The monomers, acrylic acid (AA), 2-acrylamido-2-methyl-1-propane sulfonic acid (AMPSA) (Polysciences), and *N,N*-dimethylacrylamide (DMAM) (Aldrich), were used as received. Ammonium persulphate (APS, Serva), potassium metabisulphite (KBS, Aldrich), 2-aminoethanothiol hydrochloride (AET, Aldrich) and 1-(3-(dimethylamino) propyl)-3-ethyl-carbodiimide hydrochloride (EDC, Aldrich) were used for the synthesis of the graft copolymers.

For the adjustment of the pH citric acid (CA) (Merck) was used.

Water was purified by means of a Seralpur Pro 90C apparatus combined with a USF Elga laboratory unit. For the SANS experiments, deuterium oxide (Aldrich) was used.

**Polymer synthesis and characterization.** Amine-terminated PDMAM was synthesized by free radical polymerization of DMAM in water at 30 ºC for 6 h using the redox couple APS and AET as initiator and chain transfer agent, respectively. The polymer was purified by dialysis against water through the same membrane above, and finally obtained by freeze-drying. Its number average molecular weight was determined by end group titration with NaOH after neutralization with an excess of HCl, using a Metrohm automatic titrator (model 751 GPD Titrino) and 17,000 g/mol were obtained.

A copolymer of AA and AMPSA, P(AA-*co*-AMPSA), was prepared by free radical copolymerization of the two monomers in water, after partial neutralization (90% mole) with NaOH at pH ≈ 6-7, at 30 ºC for 6 h, using the redox couple APS/KBS. The product obtained was then fully neutralized (pH=11) with an excess of NaOH, purified by dialysis against water, and received in its sodium salt form after freeze-drying. Its composition, determined by acid-base titration and elemental analysis, was 18% in AA units. Its apparent weight average molecular weight, $M_w$ = 2.7 x $10^5$ g/mol, was determined by static light scattering in 0.1 M NaCl.

The graft copolymer, P(AA-*co*-AMPSA)-*g*-PDMAM, was synthesized by a coupling reaction between P(AA-*co*-AMPSA) and amine-terminated PDMAM. The two polymers were dissolved in water at a 1:1 weight ratio. Then, an excess of the coupling agent, EDC, was added and the solution was let under stirring for 6 h at room temperature. Addition of EDC was repeated for a second time. The product was purified with a Pellicon system, equipped with a tangential flow filter membrane (Millipore, cut off = 100 kDa), and freeze dried. Its composition in PDMAM side chains was found to be equal to 48 wt% (using elemental analysis), corresponding to about 14 chains per graft copolymer. A schematic depiction of the graft copolymer is presented in Scheme 1. Its apparent molecular weight, $M_w = 4.8 \times 10^6$ g/mol, was determined by static light scattering in 0.1 M NaCl.

**Rheology.** Steady-state shear viscosity measurements of semidilute aqueous polymer mixtures were performed using a Rheometrics SR 200 controlled-stress rheometer, equipped with a cone and plate geometry (diameter = 25 mm, angle = 5.7°, truncation = 56 μm). An Anton Paar AMVn automated microviscometer, equipped with a 1.8 mm diameter glass capillary and a 1.5 mm diameter steel ball, was used to measure the viscosity of the dilute solutions. The temperature was fixed at 25±0.1 °C.

**Dynamic Light Scattering (DLS).** The intensity time correlation functions $g^{(2)}(t)$ of the polarised light scattering were measured at θ = 90° at 24 °C with a full multiple tau digital correlator (ALV-5000/FAST) with 280 channels. The excitation light source was a He-Ne laser (Melles-Griot) operating at 632.8 nm, with a stabilised power of 17 mW. The incident beam was polarised vertically with respect to the scattering plane using a Glan polariser. The scattered light from the sample was collected through a Glan-Thomson polariser (Halle, Berlin) with an extinction coefficient better than $10^{-7}$. The samples used were dust-free and optically homogeneous. The intensity time correlation functions $g^{(2)}(t)$ were analysed using the inverse Laplace transformation (ILT) method with the aid of the CONTIN code.[35] From the relaxation times obtained by the ILT analysis, the translational diffusion coefficient, $D_T$, of the particles was determined by means of the equation

$$D_T = (\tau q^2)^{-1} \qquad (1)$$

where τ is the relaxation time and q the wave vector given by q = 2πnsin(θ/2)/λ, where n is the refractive index of the medium and λ the wave length of the light beam. $D_T$ was related to the hydrodynamic radius, $R_H$, of the particles through the Stokes-Einstein equation,

$$R_H = K_B T / 6\pi\eta_0 D_0 \qquad (2)$$

where $K_B$ is the Boltzmann constant, $T$ the absolute temperature, $\eta_0$ the viscosity of the solvent and $D_0$ the translational diffusion coefficient at zero concentration.

**Static Light Scattering (SLS).** SLS measurements were conducted by means of a Model MM1 SM 200 spectrometer (Amtec, France). An He-Ne 10mW laser operating at 633 nm was used as a light source and a complete series of measurements at different angles and concentrations were conducted for molecular weight determination. The solutions used, dust-free and optically transparent, were centrifuged for two hours at 15.000 turns per minute. The refractive index increment, *dn/dC*, value was measured by means of a Chromatix KMX 16 differential refractometer operating also at 633 nm. The results obtained were subjected to a Zimm analysis, and the molecular weight of the complexes was determined as the average of the values found by extrapolation to zero angle and zero concentration.

**Small-angle neutron scattering (SANS).** SANS measurements were carried out at the Laboratoire Léon Brillouin (Saclay, France). The data were collected on beam line PACE at three configurations (6 Å, sample-to-detector distances 1 m; 7 Å and 18 Å, 4.55 m), covering a broad q range from 0.0023 to 0.32 Å$^{-1}$. 5 mm light path quartz cells were used. Empty cell scattering was subtracted and the detector was calibrated with 1 mm $H_2O$ scattering. All measurements were carried out at room temperature. Data were converted to absolute intensity through a direct beam measurement, and the incoherent background was determined with $H_2O/D_2O$ mixtures.

**Atomic Force Microscopy (AFM).** AFM images were collected under ambient conditions (23°C, 50 % RH) using tapping mode AFM (TM-AFM) on a NanoScope III multimode scanning probe microscope (Veeco, USA). Silicon tips with a spring constant of 42 N m$^{-1}$ and a resonance frequency of approximately 320 kHz were used. In tapping mode[36], the cantilever oscillates at its resonance frequency (typically 200-400 Hz in air), so that the tip interacts very briefly with the surface during each oscillation cycle with a small amplitude (A ~ 10 nm). The reduction of the cantilever oscillation from its set point value, due to interactions between the AFM tip and the sample during the scan, is used to determine the topography of the surface. In order to minimise the forces of interaction, the ratio of the set point value to the free amplitude of the cantilever was maintained at approximately 0.9 ("light tapping") by adjusting the vertical position of the sample. Images were recorded with a resolution of 512 x 512 pixels and a scan rate of 0.5-0.8 Hz. Height and lateral dimensions of the particles were measured using the Nanoscope image analysis software (NanoScope V6.13)

Samples were prepared by depositing a drop (5µL) of polymer mixture solution (3,6 × $10^{-5}$ g.cm$^{-3}$) on freshly cleaved mica. The sample was then gently allowed to evaporate under ambient conditions in a Petri Dish and observed after 20 minutes.

**Preparation of the polymer mixture solutions.** Stock solutions of the mixture G48/PAA were prepared by mixing a 5.50 x $10^{-2}$ g/cm$^3$ G48 solution with a 2.10 x $10^{-2}$ g/cm$^3$ PAA solution in D$_2$O for the SANS measurements and a 5.2 x $10^{-2}$ g/cm$^3$ G48 solution with a 2.00 x $10^{-2}$ g/cm$^3$ PAA solution in H$_2$O for all the other measurements, at pH = 2.0, adjusted with CA. The mixtures prepared were considered to contain PAA chains in equivalent quantities with the PDMAM side chains of the graft copolymer G48, that is, in a unit mole ratio PAA/PDMAM 1.1/1, according to our previous study[23]. At this point, we consider it useful to point out that a simple complex should be comprised by one G48 macromolecule and almost two PAA chains, so that its molecular weight should be of the order of 7 x $10^5$ g/mol. All dilutions were realized with 0.05M CA (pH = 2.0). The solutions after their preparation were let under agitation for 24 hours at room temperature.

**Results and Discussion**

**Rheology.** Figure 1 shows the variation of the Newtonian viscosity, η, vs. the concentration, c, for a G48/PAA mixture in an aqueous solution, at pH = 2.0. We have chosen the stoichiometric composition, corresponding to a unit mole PAA/PDMAM ratio 1.1, as determined from SANS measurements at different polymer mixture ratios in a previous work[23]. We see that a critical concentration $c^* = 7.0$ x $10^{-3}$ g/cm$^3$ appears separating the dilute from the semidilute concentration regions. In the semidilute concentration region η increases with c and follows a scaling law with an exponent equal to 7.5. This high value shows that a physical gel is formed, due to the interconnection of the anionic backbone chains of G48 in a transient network, through the hydrogen bonding interpolymer complexes formed between its PDMAM side chains and PAA.

**DLS.** Figure 2 shows the intensity time correlation functions, $g^{(2)}(t)$, and the ILT distributions for the same as above G48/PAA mixture in solution at pH = 2.0 at six different concentrations, from $0.55 \times 10^{-3}$ g/cm$^3$ to $1.8 \times 10^{-2}$ g/cm$^3$. The time correlation functions curves obtained are generally indicative of a system comprised of colloidal particles. We observe that at low concentrations, c = $0.55 \times 10^{-3}$ g/cm$^3$, $1.10 \times 10^{-3}$ and $1.7 \times 10^{-3}$ g/cm$^3$, Figures 2(a), (b) and (c) respectively, single ILT distributions, around $1 \times 10^{-3}$ s appear. At a higher concentration, c = $6.8 \times 10^{-3}$ g/cm$^3$, Figure 2(d), i.e. close to c*, a broadening of the distribution to higher times appears, indicating a slowing of the diffusion times, explained by an increase in the interactions between the particles. This behavior, that is in accordance with the viscosity behavior observed above, is even more accentuated as concentration increases further at c = $9.0 \times 10^{-3}$ g/cm$^3$, Figure 2(e), that is higher than c*, where a second peak in the distribution curve appears at about one order of magnitude higher. Finally, at c = $1.80 \times 10^{-2}$ g/cm$^3$, Figure 2(f), a third peak appears at much higher time, while the correlation function curve is not anymore indicative of any independent particles in the system. A dramatic slow down in motion occurs due to the formation of a transient network taking place in this high concentration region.

Figure 3 shows the concentration dependence of the diffusion coefficient, D, calculated by means of eq 1, for the three dilute solutions shown in Figures. 2 (a), (b) and (c). The relaxation time for each solution was obtained by the peak of the corresponding ILT distribution curve. From the value obtained by extrapolation to zero concentration, $D_0 = 2.33 \times 10^{-8}$ cm$^2$s$^{-1}$, and using eq 2, where we have put T = 277 K as the room temperature and $\eta_0$ = 0.89 cp for the viscosity of water, a value equal to 105 nm was obtained for the hydrodynamic radius $R_H$ of the particles at infinite dilution.

**SLS.** SLS measurements at different angles were performed with dilute solutions and by extrapolation to zero angle and zero concentration, according to a Zimm plot shown in Figure 4, the weight average molecular weight, $M_w$, and the radius of gyration, $R_G$, of the colloidal nanoparticles were determined. A value of $M_w$ = $5.7 \times 10^6$ g/mol was obtained for the molecular weight, somewhat higher but comparable to the value M = $4.5 \times 10^6$ g/mol, calculated after SANS measurements in the following. An aggregation number equal to 6 - 8 can be calculated on the basis of these molecular weight results, showing that each colloidal nanoparticle should be comprised by 6 – 8 graft copolymer chains and 12 – 16 PAA chains. Regarding the radius of gyration, the value $R_G$ = 85 nm was obtained, which combined with the hydrodynamic radius, $R_H$ = 105 nm, found above, shows that the colloidal particles should be of a spherical form, as $R_G/R_H$ is close to the square root of 3/5.

**SANS.** Figure 5 shows the variation of the SANS intensity, *I*, versus the scattering wave vector, *q*, for the same G48/PAA90 mixture at six different concentrations in $D_2O$, at pH = 2. The scattered intensity can be discussed separately for three different q regions. At low q, roughly q < 0.01 Å$^{-1}$, typical Guinier scattering is found at low concentration, indicative of the finite size of aggregates. At intermediate q, 0.01 < q < 0.1 Å$^{-1}$, we observe that *I* decreases abruptly, following a scaling law of the form $I \sim q^{-d}$. As the values of the exponent d vary between 3.5 and 4.0, the presence of three-dimensional objects with smooth or fractal surfaces is indicated,[37] which we attribute to the insoluble hydrogen-bonding interpolymer complexes formed between the PDMAM side chains of the graft copolymer and the PAA chains.[17, 23] At high q, q > 0.1 Å$^{-1}$, finally, chain scattering is found, which should be attributed to the anionic backbone of the graft copolymer, comprising the hydrophilic shell of the colloidal particles formed.

Furthermore, as the concentration becomes higher than $c^*$, the critical overlapping concentration, the form of the intensity curves changes, with a tendency to shift to lower values at low *q* and to exhibit a structural peak, at around 0.01 Å$^{-1}$, which becomes clear only in the most concentrated solution, Figure 3(f), reflecting the interactions between the objects. By considering that it corresponds to the most probable distance between them, we can apply a cubic lattice model based on the mass conservation of the complex particles, with the distance between the particles given through $D = 2\pi/q_0$. Since the volume, *V*, of each particle can be estimated by $V = \varphi D^3$, where $\varphi$ is the volume fraction of the particles, their "dry" radius, $R_{dry}$, can be calculated by

$$R_{dry} = \sqrt[3]{\frac{6\pi^2 \varphi}{q_0^3}} \qquad (3)$$

In the case of the most concentrated solution, Figure 5(f), where $q_0 = 0.0108$ Å$^{-1}$, and $\varphi$ equals to $1.85 \times 10^{-2}$, calculated by taking into account only the compact complex particles formed between the PDMAM side chains of G48 and the PAA chains and their mass density, $d_c = 1.28$ g/cm$^3$, determined elsewhere[23], $R_{dry}$ comes out to be equal to 9.5 nm.

The low $q$ intensity region corresponds to a first approximation to the Guinier regime of the scattering of individual non-interacting, finite-sized objects;[38] their radius leads to a characteristic decrease in I, whose magnitude is related to their mass. If the objects are spheres of radius $R_c$:

$$I = I_0 \exp(-R_c^2 q^2 / 5) \qquad (4a)$$

with

$$I_0 = \varphi \, \Delta\rho^2 \, V_0 \qquad (4b)$$

where $V_0$ denotes the dry volume of an individual object, $\varphi$ the volume fraction of the objects and $\Delta\rho$ the scattering contrast between the solvent and the dry polymer.

Then, from eq 4(b), by taking $\Delta\rho = 5.0 \times 10^{10}$ cm$^{-2}$,[23] we obtain the value of 165 cm$^{-1}$ for the intensity at zero $q$, $I_0$. Using this value in the Guinier form expressed by eq 4(a), we obtain a relatively good fit for the data of Figure 5(f), if we use a value equal to 16 nm for the radius, $R_c$, of the compact particles. We also observe that we have relatively good Guinier fitting for all the concentrations measured by using as $I_0$ the value occurring from the initially estimated quantity for the most concentrated solution, 165 cm$^{-1}$, adjusted each time proportionally to the concentration. The value for the radius of the particles obtained is practically stable at 16 – 17 nm. It should also be considered as a "wet" radius representing about 80% hydrated particles. Moreover, it should be compared to the hydrodynamic radius of the particles, $R_H$ = 105 nm, obtained from DLS measurements in dilute solution. It is noteworthy that this hydrodynamic radius includes not only the insoluble core of the compact hydrogen-bonding interpolymer complexes formed between PAA and the PDMAM side chains of the graft copolymer, but also a hydrophilic shell comprised by its anionic backbone. A representative schematic depiction of the colloidal nanoparticles formed is presented by Scheme 2. The hydrophilic shell is comprised by loops and single strands of the anionic backbone, extended, due to their charge and the low ionic strength of the solution, while their length should be related with the distribution of the PDMAM side chains in the ionic backbone and its length, estimated to be over the 330 nm on the basis of its molecular weight.

From the volume of the particle we can also obtain the molecular mass, $M_c$, of the dry complex particle

$$M_c = V \, d_c \, N_A \qquad (5)$$

where $N_A$ is Avogadro's number. Eq 5 gives $M_c = 2.8 \times 10^6$ Da, corresponding to a value equal to $4.5 \times 10^6$ Da for the whole particle. This core molecular weight value also implies that each particle contains about 90 PDMAM side chains involving more than six graft copolymer chains. This leads to the formation of a transient network explaining the increase in viscosity observed in semidilute solution, Figure 1, and the gel formation already studied.[17, 21]

**AFM.** Figure 6 represents an AFM image showing globular particles homogeneously distributed on the mica surface. The particles obtained are characterized by lateral dimensions of about 60 nm, with a height of about 1.5 – 2 nm. It is well established that lateral dimensions, determined by AFM, are overestimated due to the convolution effect of the AFM tip when scanning small objects.[39] Assuming a tip radius of 10 nm (the actual size of commercial AFM tips is given between 5 and 15 nm) we can estimate a true lateral size around 22.5 nm. From these dimensions and assuming spherical cap geometry, the particles volume deposited on the substrate is about 3200 nm$^3$. It appears that this estimated volume is lower than the one of the insoluble core in solution, as determined by SANS (17150 nm$^3$), but it is in a fairly good agreement with the dry volume of the core, as it has been estimated to be 80% hydrated. On the other hand, the drying procedure used in the preparation of the sample before AFM imaging, which is useful to immobilize the polymer on the mica substrate (both mica and the polymer are negatively charged), may induce a collapse of the colloidal nanoparticules due to a conformation change upon deposition on the mica and/or a possible dehydration of the polymer. As a consequence, the observed dimension and shape of the nanoparticles observed by AFM should look like the one of the dry core. This finding emphasizes the fact that it is a multi-scale organized particle with a central hydrophobic core, hydrated up to 80%, comprised by the hydrogen-bonding IPC formed between PAA and the PDMAM side chains of G48, and an hydrophilic shell made of the P(AA-*co*-AMPSA) anionic backbone of the G48.

**Conclusions**

We have studied the hydrogen-bonding interpolymer complexation between PAA and PDMAM grafted onto a negatively charged backbone (P(AA-*co*-AMPSA)) by viscometry, dynamic and static light scattering, as well as by SANS and AFM. The results obtained in aqueous solution at pH = 2.0 revealed a structured system consisting of anionic colloidal nanoparticles. According to dynamic and static light scattering results spherical particles are formed with a hydrodynamic radius of about 105 nm. They are comprised by a compact core of PAA/PDMAM hydrogen-bonding interpolymer complexes, and a hydrophilic shell of anionic P(AA-*co*-AMPSA) chains. SANS measurements showed that the hydrophobic core presents a radius of 16 – 17 nm and a molar mass of $2.8 \times 10^6$ Da. AFM revealed the formation of particles with a size approaching that of the hydrophobic core.

**Acknowledgment.** This research project has been supported by the European Commission under the 6th Framework Programme through the Key Action: Strengthening the European Research Area, Research Infrastructures. Contract n°: HII3-CT-2003-505925.

**Scheme 1.** A schematic depiction of the graft copolymer P(AA-*co*-AMPSA)-*g*-PDMAM (G48).

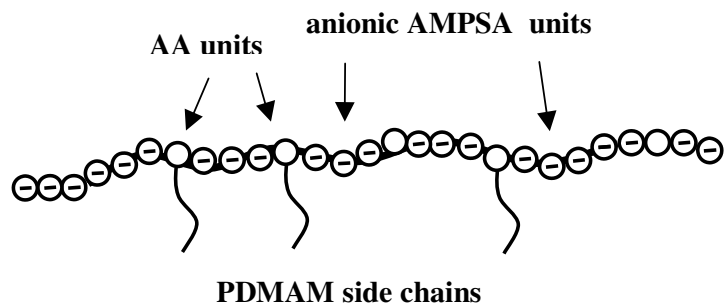

**Scheme 2.** Negatively charged colloidal particles formed through hydrogen-bonding interpolymer complexation of PAA with the PDMAM side chains of the graft copolymer P(AA-*co*-AMPSA)-*g*-PDMAM (G48), at low pH.

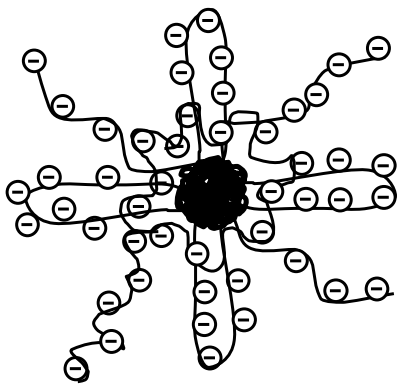

**Figure Captions**

**Figure 1.** Viscosity, η, versus concentration, c, for the polymer mixture G48/PAA in aqueous solution at pH = 2.0.

**Figure 2.** Intensity time correlation functions, $g^{(2)}(t)$, and ILT distributions for the polymer mixture G48/PAA in aqueous solution at pH = 2.0, at different concentrations: (a), c = 0.55 x $10^{-3}$ g/cm$^3$; (b), c = 1.1 x $10^{-3}$ g/cm$^3$; (c), c = 2.2 x $10^{-3}$ g/cm$^3$; (d), c = 6.8 x $10^{-3}$ g/cm$^3$; (e), c = 9.0 x $10^{-3}$ g/cm$^3$; (f), c = 1.8 x $10^{-2}$ g/cm$^3$.

**Figure 3.** Variation of the translational diffusion coefficient $D_T$ as a function of the concentration, c, in the low concentration region, for the mixture G48/PAA at pH = 2.0, and extrapolation to zero concentration.

**Figure 4.** Zimm plot for the polymer mixture G48/PAA at pH = 2.0.

**Figure 5.** SANS intensity variation vs. the wave vector q, for the G48/PAA polymer mixture in solution in D$_2$O, at pH=2, at different concentrations: (a), c = 1.6 x $10^{-3}$ g/cm$^3$; (b), c = 3.2 x $10^{-3}$ g/cm$^3$; (c), c = 6.3 x $10^{-3}$ g/cm$^3$; (d), c = 9.5 x $10^{-3}$ g/cm$^3$; (e), c = 1.9 x $10^{-2}$ g/cm$^3$; (f), c = 3.8 x $10^{-2}$ g/cm$^3$.

**Figure 6.** Tapping mode AFM picture of the G48/PAA polymer mixture deposited on mica and observed under ambient conditions.

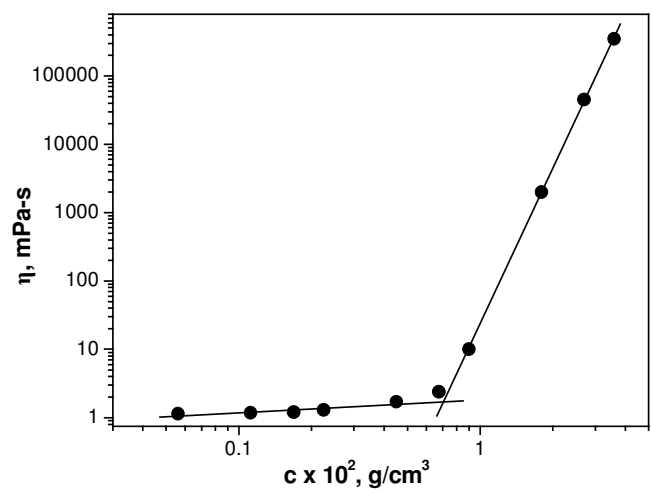

Fig. 1

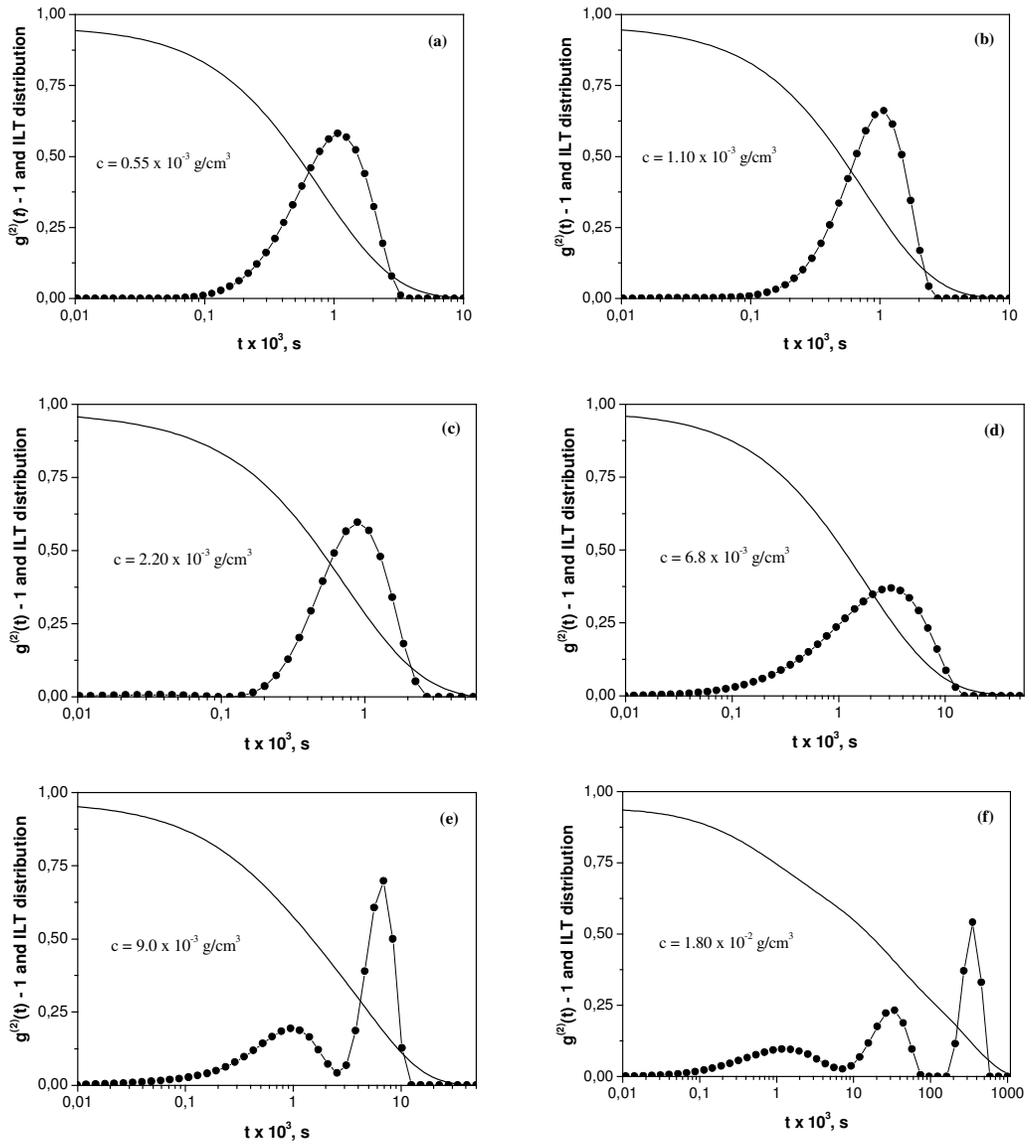

Fig. 2

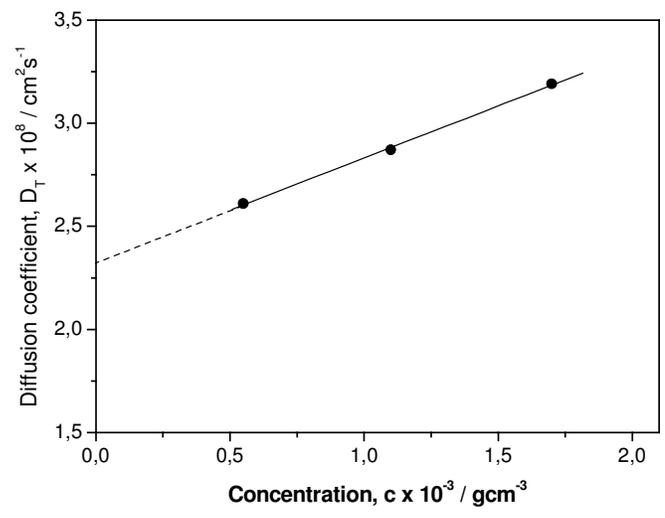

Fig. 3

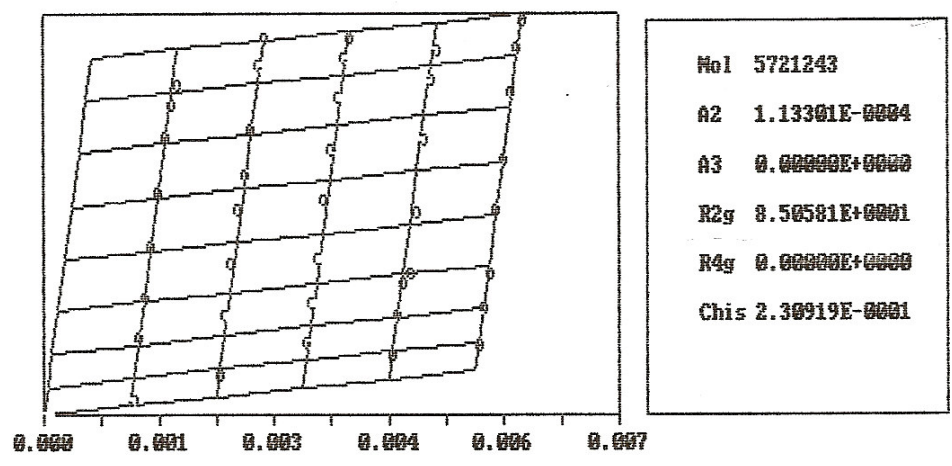

Fig. 4

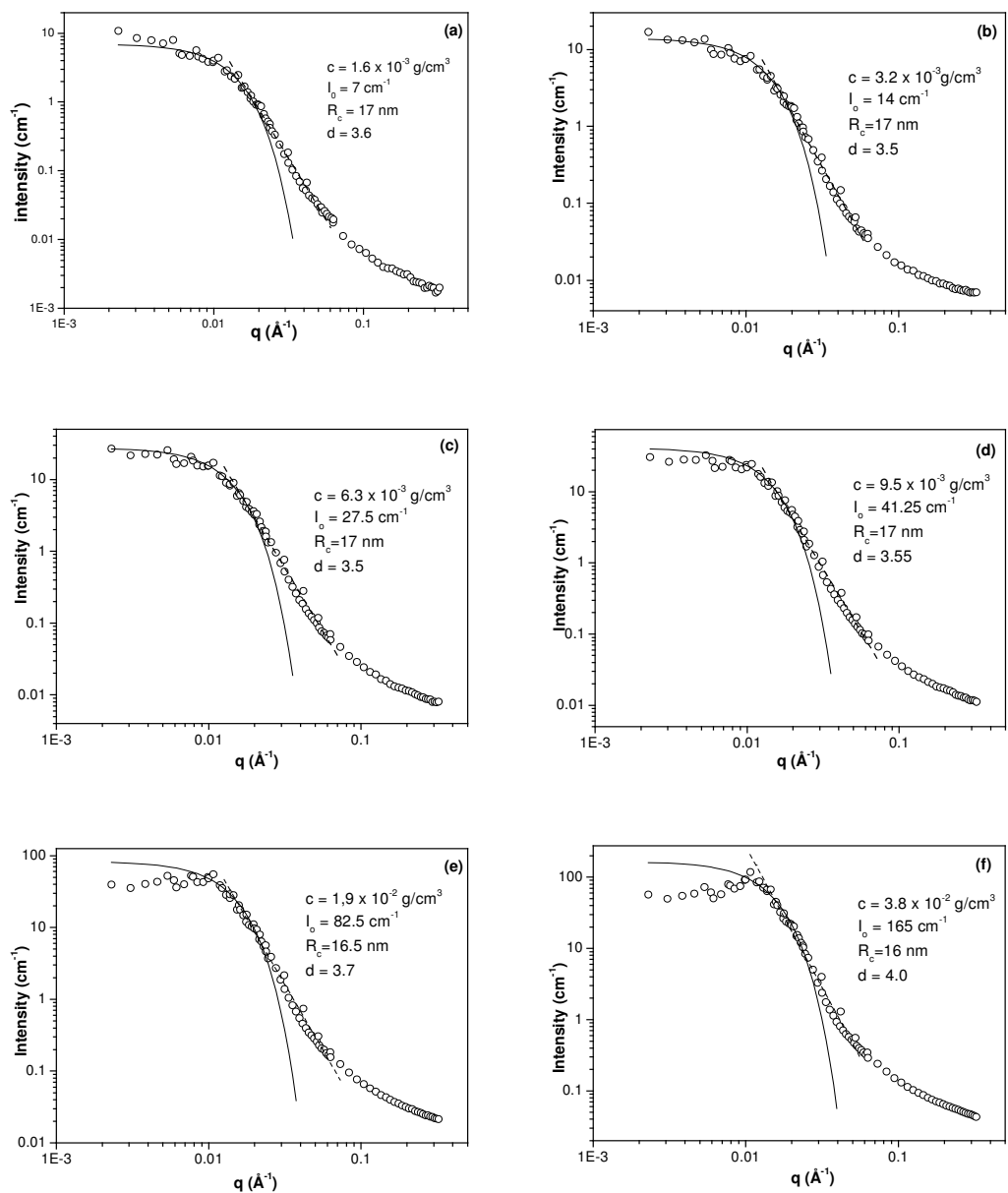

Fig. 5

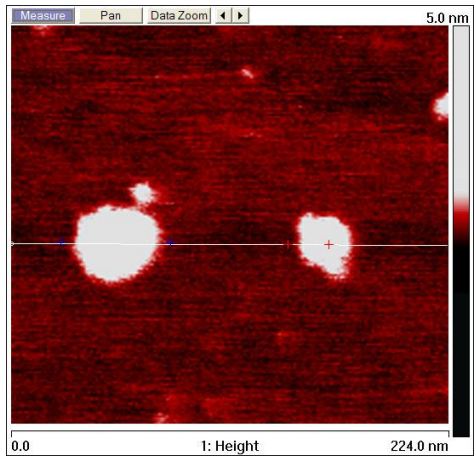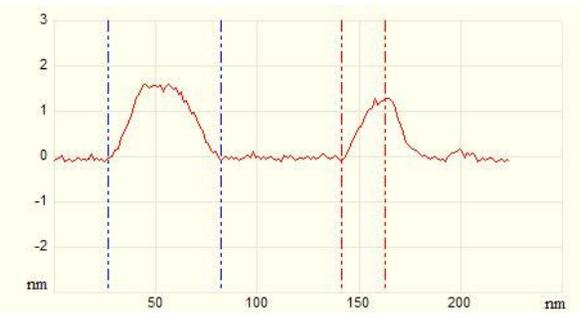

particle 1 (Z = 1,85 nm, lateral dist. = 59 nm)
particle 2 (Z = 1,5 nm, lateral dist. = 45 nm)

Fig. 6

**Table of Contents Graphic**

Characterization of the Core-Shell Nanoparticles Formed as Soluble at low pH Hydrogen-bonding Interpolymer Complexes

Maria Sotiropoulou, Frederic Bossard, Eric Balnois, Julian Oberdisse and Georgios Staikos

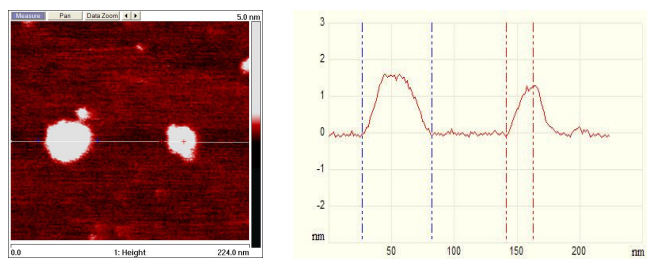

**References**

(1) Bailey, F. E.; Lundberg, Jr., R. D.; Callard, R. W. *J. Polym. Sci. Part A* **1964**, *2*, 845-851.

(2) Ikawa, T.; Abe, K.; Honda, K.; Tsuchida, E. *J. Polym. Sci.Polym. Chem. Ed.* **1975**, *13*, 1505-1514.

(3) Klenina, O. V.; Fain, E. G. *Polym. Sci. U.S.S.R.* **1981**, *23*, 1439-1446.

(4) Eustace, D. J.; Siano, D. B.; Drake, E. N. *J. Appl. Polym. Sci.* **1988**, *35*, 707-716.

(5) Khutoryanskiy, V. V.; Dubolazov, A. V.; Nurkeeva, Z. S.; Mun, G. A. *Langmuir* **2004**, *20*, 3785-3790.

(6) Aoki, T.; Kawashima, M.; Katono, H.; Sanui, K.; Ogata, N.; Okano, T.; Sakurai, Y. *Macromolecules* **1994**, *27*, 947-952.

(7) Mun, G. A.; Nurkeeva, Z. S.; Khutoryanskiy, V. V.; Sarybayeva, G. S.; Dubolazov, A. V. *Eur. Polym. J.* **2003**, *39*, 1687-1691.

(8) Bekturov, E. A.; Bimendina L. A. *Adv. Polym. Sci.* **1981**, *41*, 99-147.

(9) Tsuchida, E.; Abe, K. *Adv. Polym. Sci.* **1982**, *45*, 1-119.

(10) Ozeki, T.; Yuasa, H.; Kanaya, Y. *J. Controlled Release* **2000,** *63*, 287-295.

(11) Lele, B. S.; Hoffman, A. S. *J. Controlled Release* **2000**, *69*, 237-248.

(12) Carelli, V.; Di Colo, G.; Nannipieri, E.; Poli, B.; Serafini, M. F. *Int. J. Pharm.* **2000**, *202*, 103-112.

(13) Chun, M.-K.; Cho, C.-S.; Choi, H.-K. *J. Controlled Release*, **2002**, *81*, 327-334.

(14) Mathur, A. M.; Drescher, B.; Scranton, A. B.; Klier, J. *Nature* **1998**, *392*, 367-370.

(15) Umana, E. ; Ougizawa, T. ; Inoue, T. *J. Membrane Sci.* **1999**, *157*, 85-96.

(16) Bell, C. L.; Peppas, N. A.; *Adv. Polym. Sci.* **1995**, *122*, 125-175.

(17) Sotiropoulou, M.; Bokias, G. ; Staikos, G. *Macromolecules*, **2003,** *36*,1349-1354.

(18) Ivopoulos, P.; Sotiropoulou, M.; Bokias, G.; Staikos, G. *Langmuir,* **2006,** *22,* 9181-9186.

(19) Wang, Y.; Morawetz, H. *Macromolecules* **1989**, *22*, 164-167.

(20) Shibanuma, T.; Aoki, T.; Sanui, K.; Ogata, N.; Kikuchi, A.; Sakurai, Y.; Okano, T. *Macromolecules* **2000**, *33*, 444-450.

(21) Bossard, F.; Sotiropoulou, M.; Staikos, G. *J. Rheol.* **2004**, *48*, 927-936.

(22) Zeghal, M.; Auvray, L. *Europhys. Lett.* **1999**, *45*, 482-487.

(23) Sotiropoulou, M.; Oberdisse, J.; Staikos, G. *Macromolecules* **2006**, *39*, 3065-3070.

(24) Bronich, T. K.; Popov, A. M.; Eisenberg, A.; Kabanov, V. A.; Kabanov, A. V. *Langmuir*, **2000**, *16*, 481-489.

(25) Harada, A.; Kataoka, K. *Science*, **1999**, *283*, 65-67.

(26) Maruyama, A.; Katoh, M.; Ishihara, T.; Akaike, T. *Bioconjugate Chem.* **1997**, *8*, 3-6.

(27) van der Burgh, S.; de Kaizer, A.; Cohen Stuart, M. A.; *Langmuir* **2004,** *20,* 1073-1084.

(28) Voets, I. K.; de Kaizer, A.; Cohen Stuart, M. A.; de Waard, P. *Macromolecules* **2006,** *39,* 5952-5955.

(29) Hervé, P.; Destarac, M.; Berret, J.-F.; Lal, J.; Oberdisse, J.; Grillo, I. *Europh. Lett.* **2002**, *58*, 912-918.

(30) Berret, J.-F.; Vigolo, B.; Eng, R.; Hervé, P. *Macromolecules* **2004**, *37*, 4922-4930.

(31) Nisha, C. K.; Basak, P.; Manorama, S. V.; Maiti, S.; Jayachandran, K. N. *Langmuir* **2003**, *19*, 2947-2955.

(32) Balomenou, I.; Bokias, G. *Langmuir* **2005**, *21*, 9038-9043.

(33) Tsolakis, P.; Bokias, G. *Macromolecules* **2006**, *39*, 393-398.

(34) Staikos, G.; Karayanni, K.; Mylonas, Y. *Macromol. Chem. Phys.* **1997**, *198*, 2905-2915.

(35) (a) Provencher S. W.; *Comput. Phys. Commun.* **1982,** *27,* 213-227. (b) Provencher S. W.; *Comput. Phys. Commun.* **1982,** *27*, 229-242.

(36) Zhong, Q.; Innis, D.; Kjoller, K.; Elings, V.B. *Surf. Sci. Lett.* **1993**, *290*, 688-692.

(37) Higgins, J. S.; Benoît H. C. In *Polymers and Neutron Scattering*; Oxford Science Publications; Clarenton Press: Oxford, 1994.

(38) Lindner, P.; Zemb, Th., Eds.; *Neutrons, X-rays and Light: Scattering Methods Applied to Soft Matter*; North-Holland, Delta Series, Elsevier: Amsterdam, 2002.

(39) Westra, K.L.; Mitchell, A.W.; Thomson, D.J.; *J. Appl. Phys.* **1993,** *74*, 3608-3610.